\documentclass[conference]{IEEEtran}
\ifCLASSINFOpdf
\else
\fi
\usepackage{amsmath,dsfont,bbm,epsfig,amssymb,amsfonts,amstext,verbatim,amsopn,cite,subfigure}
\usepackage{multirow,multicol,lipsum,xfrac}
\usepackage{amsthm}
\usepackage{mathtools,amsthm}
\usepackage{url}
\usepackage{amsfonts}
\usepackage{epsfig}
\usepackage[font=small]{caption}
\usepackage{psfrag}	
\usepackage[Algorithm,ruled]{algorithm}
\usepackage{algorithmicx}
\usepackage{algpseudocode}
\usepackage{pifont}
\usepackage[utf8]{inputenc}
\usepackage[T1]{fontenc}  
\usepackage[nolist]{acronym}
\usepackage{paralist}
\usepackage{enumitem}
\usepackage{bbm}
\usepackage[process=auto]{pstool}
\usepackage{tikz,pgfplots}
\usetikzlibrary{shapes,arrows}

\newcommand{\setX}{\mathbbmss{X}}
\newcommand{\setL}{\mathbbmss{L}}

\newcommand{\setR}{\mathbbmss{R}}

\newcommand{\setW}{\mathbbmss{W}}

\newcommand{\setC}{\mathbbmss{C}}

\newcommand{\Ex}[2]{ \mathbbm{E}_{#2} \left\lbrace #1 \right\rbrace }

\newcommand{\rmj}{\mathrm{j}}

\newcommand{\out}{\mathrm{out}}
\newcommand{\In}{\mathrm{in}}

\newcommand{\argmin}{\mathop{\mathrm{argmin}}}

\newcommand{\bxx}{\mathbf{x}}
\newcommand{\mh}{\mathbf{h}}
\newcommand{\bss}{\mathbf{s}}
\newcommand{\bww}{\mathbf{w}}
\newcommand{\buu}{\mathbf{u}}
\newcommand{\bvv}{\mathbf{v}}
\newcommand{\byy}{\mathbf{y}}
\newcommand{\bgg}{\mathbf{g}}
\newcommand{\bff}{\mathbf{f}}

\newcommand{\rmw}{\mathrm{w}}
\newcommand{\rmu}{\mathrm{u}}
\newcommand{\rmv}{\mathrm{v}}
\newcommand{\rmy}{\mathrm{y}}

\newcommand{\bx}{{\boldsymbol{x}}}

\newcommand{\set}[1]{\left\lbrace#1\right\rbrace}

\newcommand{\brc}[1]{\left( #1 \right) }
\newcommand{\dbc}[1]{\left[ #1 \right] }

\newcommand{\bz}{{\boldsymbol{z}}}

\newcommand{\bs}{{\boldsymbol{s}}}

\newcommand{\bu}{{\boldsymbol{u}}}
\newcommand{\bv}{{\boldsymbol{v}}}

\newcommand{\bw}{{\boldsymbol{w}}}

\newcommand{\bzz}{{\mathbf{z}}}

\newcommand{\by}{{\boldsymbol{y}}}

\newcommand{\trp}{\mathsf{T}}

\newcommand{\mR}{\mathbf{R}}

\newcommand{\mI}{\mathbf{I}}

\newcommand{\mQ}{\mathbf{Q}}

\newcommand{\mH}{\mathbf{H}}

\newcommand{\rf}{{\mathrm{RF}}}
\newcommand{\rss}{{\mathrm{RSS}}}

\newcommand{\norm}[1]{\lVert #1 \rVert}

\newcommand{\abs}[1]{\lvert #1 \rvert}

\newcommand{\gred}[1]{\nabla_{\hspace*{-.5mm}#1} \hspace*{.5mm}}
\newcommand{\gredSq}[1]{\nabla^2_{\hspace*{-.5mm}#1} \hspace*{.5mm}}

\newtheoremstyle{mystyle}%                % Name
  {}%                                     % Space above
  {}%                                     % Space below
  {}%                                     % Body font
  {}%                                     % Indent amount
  {\bfseries}%                            % Theorem head font
  {:}%                                     % Punctuation after theorem head
  { }%                                    % Space after theorem head, ' ', or \newline
  {}%                                     % Theorem head spec (can be left empty, meaning `normal')

\theoremstyle{mystyle}

\newtheorem{definition}{Definition}

\newtheorem{remark}{Remark}

\algnewcommand\algorithmicLet{\textbf{Let}}
\algnewcommand\Let{\item[\algorithmicLet]}
\algnewcommand\algorithmicSet{\textbf{Set}}
\algnewcommand\Set{\item[\algorithmicSet]}

\algnewcommand\algorithmicInitiate{\textbf{Initiate}}
\algnewcommand\Initiate{\item[\algorithmicInitiate]}
\algnewcommand\algorithmicStart{\textbf{Begin}}
\algnewcommand\Begin{\item[\algorithmicStart]}
\algnewcommand\algorithmicEnd{\textbf{End}}
\algnewcommand\End{\item[\algorithmicEnd]}

\algnewcommand\algorithmicOutP{\textbf{Output:}}
\algnewcommand\Out{\item[\algorithmicOutP]}
\newcommand\NoDo{\renewcommand\algorithmicdo{}}

\newcounter{bar}

\begin{document}

\begin{acronym}
\acro{mimo}[MIMO]{multiple-input multiple-output}
\acro{csi}[CSI]{channel state information}
\acro{awgn}[AWGN]{additive white Gaussian noise}
\acro{iid}[i.i.d.]{independent and identically distributed}
\acro{uts}[UTs]{user terminals}
\acro{bs}[BS]{base station}
\acro{tas}[TAS]{transmit antenna selection}
\acro{glse}[GLSE]{generalized least square error}
\acro{rhs}[r.h.s.]{right hand side}
\acro{lhs}[l.h.s.]{left hand side}
\acro{wrt}[w.r.t.]{with respect to}
\acro{rs}[RS]{replica symmetry}
\acro{rsb}[RSB]{replica symmetry breaking}
\acro{papr}[PAPR]{peak-to-average power ratio}
\acro{rzf}[RZF]{regularized zero forcing}
\acro{snr}[SNR]{signal-to-noise ratio}
\acro{rf}[RF]{radio frequency}
\acro{mf}[MF]{matched filtering}
\acro{gamp}[GAMP]{generalized AMP}
\acro{amp}[AMP]{approximate message passing}
\acro{vamp}[VAMP]{vector AMP}
\acro{map}[MAP]{maximum-a-posterior}
\acro{mmse}[MMSE]{minimum mean squared error}
\acro{ap}[AP]{average power}
\acro{pa}[PA]{power amplifier}
\acro{tdd}[TDD]{time division duplexing}
\acro{rss}[RSS]{residual sum of squares}
\acro{rls}[RLS]{regularized least-squares}
\end{acronym}

\title{RLS Precoding for Massive MIMO Systems with Nonlinear Front-End}
% Authors
\author{
\IEEEauthorblockN{
Ali Bereyhi\IEEEauthorrefmark{1},
Saba Asaad\IEEEauthorrefmark{1},
Ralf R. M\"uller\IEEEauthorrefmark{1}, and
Symeon Chatzinotas\IEEEauthorrefmark{2}
}
\IEEEauthorblockA{
\IEEEauthorrefmark{1}Institute for Digital Communications, Friedrich-Alexander Universit\"at Erlangen-N\"urnberg,\\
\IEEEauthorrefmark{2}Interdisciplinary Center for Security, Reliability and Trust, University of Luxembourg,\\
\{ali.bereyhi, saba.asaad, ralf.r.mueller\}@fau.de, symeon.chatzinotas@uni.lu
\thanks{This work has been accepted for presentation in the 20th IEEE International Workshop on Signal Processing Advances in Wireless Communications (SPAWC) 2019 in Cannes, France. The link to the final version~in~the~Proceedings of SPAWC will be available later.}
}
}


\IEEEoverridecommandlockouts

\maketitle

\begin{abstract}
To keep massive MIMO systems cost-efficient, power amplifiers with rather small output dynamic ranges are employed. They may distort the transmit signal and degrade the performance. This paper proposes a \textit{distortion aware} precoding scheme for realistic scenarios in which~RF chains have nonlinear characteristics. The proposed~scheme~utilizes the method~of~regularized least-squares (RLS) to jointly compensate the channel impacts and the distortion imposed by the RF chains. 

To construct the designed transmit waveform with low computational complexity, an iterative algorithm based on approximate message passing is developed. This algorithm is shown to track the achievable average signal distortion of the proposed scheme tightly, even for practical system dimensions. The results demonstrate considerable enhancement compared to the state of the~art.
\end{abstract}
\vspace*{-3mm}

\IEEEpeerreviewmaketitle
\vspace*{-3mm}
\section{Introduction}
\label{sec:intro}
Theoretical analyses depict that linear~multiuser~\ac{mimo} precoding techniques are efficient in the large-system limit \cite{marzetta2010noncooperative}. This result, along with~the low complexity of these schemes, has introduced linear precoding as the dominant approach for signal pre-processing in massive \ac{mimo} \cite{hoydis2013massive}. Investigations in this respect however rely on~the characteristics of transceiver components which~are often described by simplified models. An exemplary component, which is the focus of this paper, is the \textit{\ac{pa}} used in the transmit \ac{rf} chains. 

For sake of simplicity, \ac{pa}s are often treated~as~linear~components. The linear model, however, is not valid in general. In fact, such a characterization is a rather good \textit{approximation}~when~the \ac{papr} of the signal~is less than the input \textit{back-off} of the \ac{pa}. Such a constraint can be easily violated in massive \ac{mimo} systems. In fact, the expense of a \ac{pa} is proportional to its linearity~characteristics. To keep massive \ac{mimo} settings cost-efficient, \ac{rf}~chains are implemented via \ac{pa}s with low back-offs. This increases the nonlinear distortion on the \ac{rf} transmit signal and degrades~the performance.

%\subsection*{Distortion-Aware Precoding}
The nonlinear distortion caused by \ac{pa}s can be effectively~resolved via~signal pre-processing. In this respect, one can~pre- distort the transmit waveform, either on the \textit{sample} or \textit{symbol} basis, such that the receive signal will be of the desired shape. For given characteristics of \ac{pa}s, this approach is analytically tractable with standard signal processing techniques, e.g. \cite{piazza2014data}. In multiuser \ac{mimo} settings, the transmit signal can be directly pre-distorted at the precoding stage. To this end, one~can~include the \ac{rf} chains in the channel model, and design the precoder for the end-to-end channel. An example of this approach was studied in \cite{spano2017symbol} where the authors developed a \textit{symbol-level} precoding scheme \cite{alodeh2018symbol} for \ac{mimo} settings with nonlinear~\ac{pa}s. Another example is \cite{spano2018faster} in which the end-to-end precoding was studied in sample domain considering waveform optimization and pulse shaping. Despite the promising performance, this approach often results in high computational complexity, due to the \textit{nonlinearity} of the end-to-end channel. 

\subsection*{Contributions}
Regardless of the channel model, precoding is effectively addressed via the method of \ac{rls}. For the classic linear model, \ac{rls} leads to \ac{glse} scheme introduced~and analyzed in \cite{bereyhi2018glse,bereyhi2017nonlinear,bereyhi2017asymptotics}. This paper extends this \ac{rls}-based methodology to \ac{mimo} settings with nonlinear \ac{rf} front-ends. To this end, a multiuser \ac{mimo} setting is considered in which the transmit \ac{rf} chains have a \textit{generic} input-output characteristic.~For~this setting,~end-to-end precoding is addressed by the \ac{rls} method considering a general set of constraints on the transmit signal.

The computational complexity of such an approach may be rather high for a large \ac{mimo} system, if it is implemented in a straightforward manner. To address this issue, we develop~an algorithm based on \ac{amp}~\cite{rangan2011generalized}. The complexity of this algorithm scales linearly with the~number of transmit antennas, and it tightly tracks the large-system performance of the proposed scheme for practical dimensions.

\subsection*{Notations}
Throughout the paper, scalars, vectors and matrices~are~represented with non-bold, bold lower case and bold upper case letters, respectively. $\mI_K$ is a $K \times K$ identity matrix,~and~$\mH^{\trp}$~is the transpose of $\mH$. The real axis and the complex~plane~are denoted by $\setR$ and $\setC$, respectively. For $s\in\setC$,~$\Re\set{s}$,~$\Im\set{s}$ and $\bss\coloneqq \left[ \Re\set{s}, \Im\set{s} \right]^\trp$ are the real part, imaginary part and augmented vector, respectively. $\Ex{\cdot}{}$ is the statistical expectation. For simplicity, $\set{1, \ldots , N}$~is abbreviated by $[N]$. For any differentiable function $\bff(\bx)=\left[ f_1(\bx), \ldots, f_n(\bx) \right]^\trp$,~the~gradient is defined as $\gred{\bx} \bff(\bx) \coloneqq [ \gred{\bx} f_1(\bx), \ldots,\gred{\bx} f_n(\bx) ]^\trp$. 

\section{Problem Formulation}
\label{sec:sys}
We consider downlink transmission in a Gaussian broadcast \ac{mimo} channel with a single \ac{bs} and $K$ single-antenna users. The \ac{bs} is equipped with an antenna array of size $M$, $L\leq M$ nonlinear \ac{rf} chains, and a switching network which connects each subset of $L$ antennas to the \ac{rf} chains. 

In the $n$-th transmission interval, the \ac{bs} intends~to~transmit data symbols ${s_1\dbc{n}, \ldots,s_k\dbc{n}}$ to the \ac{uts}. To this end, it constructs the transmit signal $\bx\dbc{n}\in\setX^{M}$~with $L$ non-zero entries via~a~precoding scheme. Here, $\setX\subseteq \setC$ is the \textit{precoding support} and contains all possible points which can be selected as constellation points by the transceiver.~For instance, in the case of per-antenna constant envelope precoding, the precoding support is
%\begin{align}
$\setX = \set{ x \in \setC: \abs{x}^2 = P}$.
%\end{align}

The $m$-th entry of the transmit signal represents the symbol which is intended to be sent over antenna element $m$. Hence, the indices of non-zero entries in $\bx\dbc{n}$ correspond to~those~antennas which are set active in transmission~time~interval~$n$. Let $\setL \dbc{n} \subseteq \dbc{M}$ denote the index set of non-zero entries in the transmit signal $\bx\dbc{n}$. The switching network connects~the~\ac{rf} chains to the antennas indexed by $\setL \dbc{n}$. The precoded signal is then transmitted via the \ac{rf} chains. %over the active antennas.

The system operates in the \ac{tdd} mode which means that the uplink and downlink channels~are reciprocal. It is assumed that the \ac{csi} is estimated at both the transmit and receive sides at the beginning of each coherence interval~within~an~estimation loop whose duration is much shorter than the coherence interval. Hence, the \ac{bs} knows the \ac{csi} prior to transmission.

\subsection{Nonlinear RF Chains}
The main component of an \ac{rf} chain is the \ac{pa} which~has nonlinear input-output characteristics, in general. Several examples of nonlinear \ac{pa} models can be followed in the literature; e.g.~\cite{saleh1981frequency,o2009new}.~For sake of generality, we consider a generic input-output characteristic for the \ac{rf} chains: Let $x$ be the symbol which is fed to an \ac{rf} chain. The output of the \ac{rf} chain is given~by $w = f_\rf \brc{x}$ with $f_\rf \brc{\cdot}: \setX \mapsto \setW$ for some $\setW\subseteq \setC$. $\setW$ describes the set of all possible constellation points after being distorted by the \ac{rf} chain. We refer to $f_\rf \brc{\cdot}$ as the \textit{\ac{rf}~conversion~function}. This function is considered to be of a general form describing various nonlinear \ac{pa} models, e.g. the well-known amplitude-to-amplitude and amplitude-to-phase distortion model. Noting that the output of an \ac{rf} chain, which is not fed by any signal, is zero, we have %the only constraint on $f_\rf \brc{\cdot}$ is %
%\begin{align}
$f_\rf \brc{0} = 0$.
%\end{align}
% \footnote{See for example the discussions in \cite{spano2017symbol} and the references therein.}

Considering the \ac{rf} conversion model, the signal entry that is observed on an active  antenna reads %
%\begin{align}
$w_m = f_\rf \brc{x_m}$,
%\end{align}
where $m\in\setL\dbc{n}$. For passive antennas,~we~further~have 
\begin{align}
w_m = 0 = f_\rf \brc{0} = f_\rf \brc{x_m},
\end{align}
where $m\in \dbc{M} \setminus \setL\dbc{n}$. As a result, we~can~compactly~represent the signal on the transmit antennas as
%\begin{align}
$\bw \dbc{n} = f_\rf \brc{ \bx \dbc{n} }$. %
%\end{align}
%where $f_\rf \brc{\cdot}$ performs entry-wise on $\bx\dbc{n}$.
To distinguish~between $\bx\dbc{n}$ and $\bw\dbc{n}$, we refer to $\bw\dbc{n}$ as the \textit{\ac{rf} transmit signal} in the transmission interval $n$.

\begin{remark}
\label{remark:1}
The \ac{rf} conversion function is often derived~via interpolating methods. %\footnote{For instance, Saleh's model is derived by parameterized curve fitting. The parameters are calculated for a given hardware via the least-squares method; see for example \cite{saleh1981frequency,o2009new} for detailed discussions.}. 
Hence, actual outputs~slightly deviate from $f_\rf \brc{x_m\dbc{n}}$. As a result, one can write
\begin{align*}
\abs{w_m \dbc{n} - f_\rf \brc{x_m\dbc{n}}}^2 \leq \epsilon
\end{align*} 
where $\epsilon \downarrow 0$ in the ideal case.%; see for example \cite{o2009new}.
\end{remark}

\subsection{Channel Model}
The \ac{rf} transmit signal is sent to the \ac{uts} over a Gaussian broadcast \ac{mimo} channel which experiences~quasi-static~fading. The receive signal at user $k$ for $n$-th interval is hence %given by
\begin{align}
y_k \dbc{n} = \left. \sqrt{\beta_k} \right. \bgg_k^\trp \bw\dbc{n} + z_k \dbc{n}
\end{align}
for $k\in \dbc{K}$. Here, $\bgg_k \in \setC^{M}$ contains the fading coefficients of the uplink channel between user $k$ and the \ac{bs}. Moreover, $\beta_k$ describes the path-loss and shadowing in the channel which is the same for all antenna elements at the \ac{bs}. The random variable $z_k \dbc{n}$ denotes \ac{awgn} and is assumed to be zero-mean with variance $\sigma^2$.

Considering the channel model, the vector of receive signals at the \ac{uts} in transmission interval $n$, i.e. $\by\dbc{n} =[y_1\dbc{n},\ldots,$ $y_K\dbc{n}]^\trp$, is compactly represented as
\begin{align}
\by \dbc{n} = \mH^\trp \bw\dbc{n} + \bz \dbc{n}
\end{align}
where $\bz\dbc{n} =[z_1\dbc{n},\ldots,z_K\dbc{n}]^\trp$ is the noise vector and %$\mH = \mG \mA$ with $\mA$ being a ${K\times K}$ diagonal matrix defined as
%\begin{align}
%\mA = \Diag{\sqrt{\beta_1}, \ldots,\sqrt{\beta_K} }.
%\end{align}
%and $\mG \in \setC^{M\times K}$ being
\begin{align}
\mH = \dbc{\mh_1, \ldots,\mh_K }
\end{align}
with $\mh_k = \left. \sqrt{\beta_k} \right. \bgg_k$ represents the uplink channel vector.

\section{RLS-Based Precoding Scheme}
\label{sec:RLS}
%In the setting, 
%Assume that we 
%There are multiple sources which distort the transmit signal $\bx$, namely noise, channel fading and the conversion at the \ac{rf} chains. As noise is an independent~random process, it is not possible to be canceled out by any post- or pre-processing of the transmit or receive signal. However, the characteristics of the two other sources are known to~the~\ac{bs}. 
%Our main goal is to design a precoding scheme which compensates signal distortions prior to transmission. To this end, we note that~t
The~ultimate aim of precoding is to pre-process signals, such that data can be recovered from the receive signal at \ac{uts} with minimal post-processing. This means that the vector of noise-free receive signals, i.e. $\mH^\trp \bw\dbc{n}$, is desired to be close to the data vector, i.e. $\bs\dbc{n} = [s_1\dbc{n},\ldots,s_K\dbc{n}]^\trp$. To formulate this interpretation of precoding, consider the following definition:

\begin{definition}[RSS]
\label{def:RSS}
Let $\bs$ be a data vector and $\bv$ represent its corresponding \ac{rf} transmit signal. For a given scaling factor~$\rho$, the \ac{rss} at the \ac{uts} is defined as
\begin{align}
\rss \brc{\bv \vert \rho, \bs, \mH} = \norm{ \mH^\trp \bv  - \left. \sqrt{\rho} \right. \bs  }^2.
\end{align}
\end{definition}

The \ac{rss} determines the squared of the Euclidean distance between the noise-free receive signals and data symbols. Using this definition, the precoding problem is interpreted as
\begin{align}
\bx\dbc{n} = \argmin_{\bu \in \setX^M} \ &\rss \brc{f_\rf \brc{\bu} \vert \rho, \bs\dbc{n}, \mH} \text{ s.t. } \ \mathcal{C} \brc{\bu} \label{eq:Optim} 
\end{align}
for some $\rho$. In \eqref{eq:Optim}, $\mathcal{C}\brc\bu$ denotes the signal constraints required to be satisfied by the transmit signal. For example, when~the number of the \ac{rf} chains $L$ is less than $M$, $\mathcal{C}\brc\bu$ includes the sparsity constraint $\norm{\bu}_0 \leq L$. The formulation in~\eqref{eq:Optim}~describes the \ac{rls} method which we discuss in the following sections. For simplicity, we drop the time index $n$ in the sequel.

\subsection{GLSE Precoding at the RF Stage}
%The precoding problem in \eqref{eq:Optim} can be directly solved in the \ac{rf} stage. This means that one can find 
The \ac{rf} transmit signal $\bw$ is directly found by solving
\begin{align}
\bw = \argmin_{\bv \in \setW^M} \ &\rss \brc{\bv \vert \rho, \bs , \mH} \text{ s.t. } \ \tilde{\mathcal{C} } \brc{\bv} \label{eq:RFOptim}
\end{align}
where $\tilde{\mathcal{C} } \brc{\bv}$ contains the signal constraints in $\mathcal{C}(\bu)$ projected on $\setW^M$ with respect to the \ac{rf} conversion function. Following the \ac{rls} method, this problem is equivalently solved by
\begin{align}
\bw = \argmin_{\bv \in \setW^M} \ \norm{ \mH^\trp \bv  - \left. \sqrt{\rho} \right. \bs  }^2 + c\brc{\bv} \label{eq:GLSE} 
\end{align}
for some penalty $c\brc{\cdot}$. The precoder in \eqref{eq:GLSE} recovers the \ac{glse} precoding scheme~at~the \ac{rf} stage \cite{bereyhi2018glse,bereyhi2017asymptotics,bereyhi2017nonlinear}. The key differences here are:
\begin{inparaenum}
\item The penalty $c\brc{\bv}$ is chosen with respect to the \ac{rf} conversion function $f_\rf\brc\cdot$.
\item The entries of the \ac{rf} transmit signal should be projected back on the precoding support $\setX$. 
\end{inparaenum}

As~the~\ac{rf}~conversion function is known to the \ac{bs}, the first task is tractable. However, the backward projection of the \ac{rf} transmit entries to the precoding support cannot be uniquely done, as the $f_\rf\brc\cdot$ is not necessarily bijective.

\subsection{Projecting RF Signals on the Precoding Support}
\label{sec:projection}
\begin{figure}[t]
\centering
\definecolor{mycolor1}{rgb}{0.00000,0.44700,0.74100}%
\begin{tikzpicture}

\begin{axis}[%
width=2in,
height=1.2in,
at={(1.978in,1.234in)},
scale only axis,
xmin=-10,
xmax=6,
xtick={-10,-4.6,3.65,6},
xticklabels={{$-10$},{$-4.6$},{$3.65$},{$6$}},
xlabel style={font=\color{white!15!black}},
xlabel={$\log \left. \abs{x_m} \right.$ in [dB]},
ymin=-7.2,
ymax=0.4,
ytick={-7,-2,0},
yticklabels={{$-7$},{$-2$},{$0$}},
ylabel style={font=\color{white!15!black}},
ylabel={$\log \left. \abs{w_m} \right.$ in [dB]},
axis background/.style={fill=white}
]
\addplot [color=mycolor1, line width=1.0pt, forget plot]
  table[row sep=crcr]{%
-10	-6.84777250166283\\
-9.9	-6.75029185494824\\
-9.8	-6.65292837621\\
-9.7	-6.55568743845375\\
-9.6	-6.45857465384088\\
-9.5	-6.36159588364676\\
-9.4	-6.26475724856788\\
-9.3	-6.16806513938392\\
-9.2	-6.07152622797953\\
-9.1	-5.97514747873057\\
-9	-5.87893616025805\\
-8.9	-5.78289985755256\\
-8.8	-5.68704648447056\\
-8.7	-5.59138429660294\\
-8.6	-5.49592190451457\\
-8.5	-5.40066828735242\\
-8.4	-5.30563280681764\\
-8.3	-5.21082522149563\\
-8.2	-5.11625570153553\\
-8.1	-5.02193484366858\\
-8	-4.92787368655216\\
-7.9	-4.83408372642342\\
-7.8	-4.74057693304373\\
-7.7	-4.64736576591141\\
-7.6	-4.55446319071694\\
-7.5	-4.46188269601099\\
-7.4	-4.36963831005125\\
-7.3	-4.27774461778957\\
-7.2	-4.18621677795644\\
-7.1	-4.09507054019405\\
-7	-4.0043222621844\\
-6.9	-3.91398892671259\\
-6.8	-3.82408815859942\\
-6.7	-3.73463824143102\\
-6.6	-3.64565813400621\\
-6.5	-3.55716748641534\\
-6.4	-3.46918665565674\\
-6.3	-3.3817367206893\\
-6.2	-3.29483949681171\\
-6.1	-3.20851754925075\\
-6	-3.12279420583255\\
-5.9	-3.03769356860271\\
-5.8	-2.95324052425219\\
-5.7	-2.86946075319796\\
-5.6	-2.78638073715878\\
-5.5	-2.70402776505879\\
-5.4	-2.62242993708351\\
-5.3	-2.54161616670606\\
-5.2	-2.46161618049448\\
-5.1	-2.38246051550549\\
-5	-2.30418051406491\\
-4.9	-2.22680831573128\\
-4.8	-2.15037684623696\\
-4.7	-2.07491980319947\\
-4.6	-2.00047163839726\\
-4.5	-1.92706753640604\\
-4.4	-1.85474338939713\\
-4.3	-1.78353576790609\\
-4.2	-1.71348188738953\\
-4.1	-1.64461957040068\\
-4	-1.57698720422927\\
-3.9	-1.51062369387002\\
-3.8	-1.44556841020553\\
-3.7	-1.38186113331441\\
-3.6	-1.3195419908441\\
-3.5	-1.25865139141926\\
-3.4	-1.19922995309248\\
-3.3	-1.14131842688201\\
-3.2	-1.08495761548336\\
-3.1	-1.03018828728608\\
-3	-0.977051085874168\\
-2.9	-0.925586435238302\\
-2.8	-0.875834440979015\\
-2.7	-0.827834787832553\\
-2.6	-0.781626633904122\\
-2.5	-0.73724850204615\\
-2.4	-0.694738168871229\\
-2.3	-0.654132551939746\\
-2.2	-0.615467595709982\\
-2.1	-0.578778156882794\\
-2	-0.544097889812928\\
-1.9	-0.511459132693711\\
-1.8	-0.480892795250402\\
-1.7	-0.452428248699083\\
-1.6	-0.426093218741832\\
-1.5	-0.401913682374434\\
-1.4	-0.379913769279522\\
-1.3	-0.360115668565329\\
-1.2	-0.342539541588087\\
-1.1	-0.327203441564233\\
-1	-0.314123240637337\\
-0.9	-0.303312565014189\\
-0.800000000000001	-0.294782738725239\\
-0.7	-0.288542736497449\\
-0.600000000000001	-0.284599146153144\\
-0.5	-0.282956140867922\\
-0.4	-0.283615461535024\\
-0.300000000000001	-0.286576409394225\\
-0.2	-0.291835848991457\\
-0.100000000000001	-0.299388221442561\\
0	-0.309225567882162\\
0.0999999999999996	-0.321337562888018\\
0.199999999999999	-0.335711557583887\\
0.3	-0.352332632040988\\
0.399999999999999	-0.371183656521039\\
0.5	-0.392245361033399\\
0.6	-0.415496412616171\\
0.7	-0.440913499696828\\
0.8	-0.468471422842651\\
0.899999999999999	-0.498143191175421\\
1	-0.529900123698459\\
1.1	-0.563711954767386\\
1.2	-0.599546942928696\\
1.3	-0.637371982351967\\
1.4	-0.677152716091931\\
1.5	-0.718853650434985\\
1.6	-0.762438269610429\\
1.7	-0.807869150178823\\
1.8	-0.85510807444773\\
1.9	-0.90411614230763\\
2	-0.954853880927225\\
2.1	-1.00728135179669\\
2.2	-1.06135825465888\\
2.3	-1.11704402792109\\
2.4	-1.17429794519303\\
2.5	-1.23307920764957\\
2.6	-1.29334703196842\\
2.7	-1.35506073364335\\
2.8	-1.41817980552169\\
2.9	-1.48266399146081\\
3	-1.54847335504102\\
3.1	-1.61556834331241\\
3.2	-1.68390984558989\\
3.3	-1.75345924734389\\
3.4	-1.82417847926411\\
3.5	-1.89603006160024\\
3.6	-1.96897714390654\\
3.7	-2.0429835403372\\
3.8	-2.11801376065567\\
3.9	-2.19403303713533\\
4	-2.271007347539\\
4.1	-2.34890343437365\\
4.2	-2.42768882062189\\
4.3	-2.5073318221556\\
4.4	-2.58780155703855\\
4.5	-2.66906795192438\\
4.6	-2.75110174575493\\
4.7	-2.83387449096001\\
4.8	-2.91735855235579\\
4.9	-3.00152710393344\\
5	-3.08635412372306\\
5.1	-3.17181438691154\\
5.2	-3.25788345738483\\
5.3	-3.34453767785795\\
5.4	-3.43175415874715\\
5.5	-3.51951076593088\\
5.6	-3.6077861075374\\
5.7	-3.69655951988867\\
5.8	-3.78581105272158\\
5.9	-3.87552145379956\\
6	-3.96567215301944\\
6.1	-4.0562452461105\\
6.2	-4.14722347801538\\
6.3	-4.23859022603497\\
6.4	-4.33032948281249\\
6.5	-4.4224258392255\\
6.6	-4.51486446724794\\
6.7	-4.60763110283874\\
6.8	-4.70071202890745\\
6.9	-4.79409405840251\\
7	-4.8877645175624\\
7.1	-4.98171122936566\\
7.2	-5.07592249721109\\
7.3	-5.17038708885581\\
7.4	-5.26509422063483\\
7.5	-5.36003354198259\\
7.6	-5.45519512027368\\
7.7	-5.55056942599709\\
7.8	-5.64614731827572\\
7.9	-5.74192003074045\\
8	-5.83787915776582\\
8.1	-5.93401664107263\\
8.2	-6.03032475670062\\
8.3	-6.12679610235306\\
8.4	-6.22342358511356\\
8.5	-6.32020040953407\\
8.6	-6.41712006609191\\
8.7	-6.51417632001276\\
8.8	-6.61136320045547\\
8.9	-6.70867499005392\\
9	-6.80610621481046\\
9.1	-6.90365163433482\\
9.2	-7.0013062324219\\
9.3	-7.09906520796152\\
9.4	-7.19692396617278\\
9.5	-7.29487811015539\\
9.6	-7.39292343275017\\
9.7	-7.49105590870081\\
9.8	-7.58927168710856\\
9.9	-7.68756708417194\\
10	-7.785938576203\\
};
\addplot [color=black, dashed, forget plot]
  table[row sep=crcr]{%
-10	-2\\
3.65	-2\\
};
\addplot [color=black, draw=none, mark size=5pt, mark=star, mark options={solid, black, thick}, forget plot]
  table[row sep=crcr]{%
-4.6	-2\\
3.65	-2\\
};
\addplot [color=black, dashed, forget plot]
  table[row sep=crcr]{%
3.65	-8\\
3.65	-2\\
};
\addplot [color=black, dashed, forget plot]
  table[row sep=crcr]{%
-4.6	-8\\
-4.6	-2\\
};
\end{axis}
\end{tikzpicture}%
\caption{An \ac{rf} conversion function given by Saleh's model with $\alpha=2.092$ and $\beta=1.247$ \cite{saleh1981frequency}.\vspace*{-6mm}}
\label{fig:1}
\end{figure}
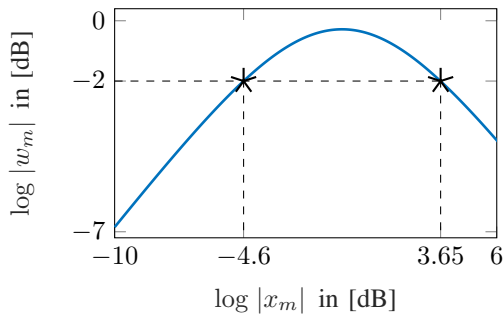

Consider the precoded signal at the \ac{rf} stage,~i.e.~$\bw$.~To~pro-ject back the \ac{rf} transmit entries on the precoding support, we need to solve %the equation
%\begin{align}
$w_m = f_\rf \brc{x_m}$ %\label{eq:Reverse}
%\end{align}
for $m\in\dbc{M}$. Considering the typical characteristics of \ac{pa}s, it is clear that there are multiple solutions for $x_m$. We hence need to select a desired solution among the available ones with respect to a metric.~To~clarify this latter statement, let us consider a simple \ac{rf} conversion function. In Fig.~\ref{fig:1}, the output amplitude, i.e. $\abs{w_m}$ of a sample \ac{pa} is sketched against the amplitude of input symbol $x_m$ using Saleh's model in \cite{saleh1981frequency} with $\alpha=2.0922$ and $\beta=1.2466$.~As~the figure depicts, $\log \abs{w_m} = -2$ dB is achieved at the output of the \ac{pa}, when the input amplitude is set either $\log \abs{x_m} = -4.6$ dB or $\log \abs{x_m} = 3.65$ dB. This behavior comes from the significant nonlinearity of the \ac{pa} in the saturation region. As we wish to restrict the power consumed in the system, we set the input symbol to the one whose amplitude is smaller, i.e., we set $\log \abs{x_m} = -4.6$ dB.

For a generic $f_\rf\brc\cdot$, the backward projection of~\ac{rf}~transmit entries on the precoding support can be formulated via the \ac{rls} method. Considering the pre-amplifier power constraint as the selection measure, the approach for choosing the input symbol in the given example is formulated as
\begin{align}
x_m = \argmin_{u\in\setX} \left. \abs{u}^2 \right. \ \text{ s.t. } \ w_m = f_\rf\brc{u}. \label{eq:RevRF1}
\end{align}
Following Remark~\ref{remark:1}, one can reformulate \eqref{eq:RevRF1} as
\begin{align}
x_m = \argmin_{u\in\setX} \left. \abs{u}^2 \right. \ \text{ s.t. } \ \abs{w_m - f_\rf\brc{u}}^2 \leq \epsilon. \label{eq:RevRF2}
\end{align}
The approach in \eqref{eq:RevRF2} is equivalently given in form~of~a~scalar \ac{rls} problem. The regularization term in this~problem~is~quad-ratic, as the metric for choosing the input symbol is its power. Nevertheless, in a generic case, the metric can be of a different form taking into account also the phase of the input symbol.

Following the above discussion, the projection approach in \eqref{eq:RevRF2} is represented in a general form as
\begin{align}
x_m = \argmin_{u\in\setX} \left. \abs{w_m - f_\rf\brc{u}}^2 + \theta \right. r\brc{u}  \label{eq:RevRF3}
\end{align}
where $r\brc{\cdot}$ is a regularization term describing the metric. For instance, in the case that we choose the transmit symbol based on its power, we have $r\brc{u} = \abs{u}^2$. $\theta$ is a tunable factor which controls the consumed power. %whose value is proportional to $\epsilon$.

\begin{remark}
Note that the signal power before and after amplification defines two different parameters which are implicitly related. The power of transmit entries, i.e.~signal~before amplification, specifies the power consumed in the system while the power at the output of transmit \ac{rf} chains mainly quantifies the multiuser interference at each UT. The \ac{rls} approach in \eqref{eq:RevRF3}, along with the \ac{glse} scheme in \eqref{eq:GLSE}, guarantees that both of these parameters are minimized given the set of available constraints on the transmit signal. In this respect, $\theta$ can be seen as a Lagrange multiplier which tunes the trade-off between the consumed power and the \ac{rss}.
\end{remark}

\section{Algorithmic Implementation via AMP}
The computational complexity of the proposed scheme is dominated by the precoding in the \ac{rf} stage. In fact,~the~projection on the precoding support deals with $M$ parallel copies of a scalar optimization problem which is addressed tractably. The calculation of the \ac{rf} transmit signal, however, requires to solve an optimization problem of size $M$ within each transmission interval\footnote{In practice, the update rate can be reduced to once per coherence~time~interval by block-wise precoding; see \cite{bereyhi2018glse}.}. In massive \ac{mimo} settings with large antenna arrays, such a task is not computationally tractable~in~practice.~To~address this issue, we propose an iterative algorithm based on \ac{amp} whose complexity scales linearly with $M$.

\subsection{RF Stage Precoding via AMP}
The \ac{glse} precoding in \eqref{eq:GLSE} is mathematically equivalent to a \textit{max-sum} problem in the Bayesian framework. This equivalence was studied in \cite{bereyhi2018precoding} where an iterative algorithm~based on \ac{gamp} \cite{rangan2011generalized,rangan2016fixed} was proposed. This algorithm is adapted to the problem in \eqref{eq:GLSE} in Algorithm~\ref{A-GAMP}.
\begin{algorithm}[t]
\caption{AMP-based Precoding in the RF Stage}
\label{A-GAMP}
\begin{algorithmic}[0]
\Initiate For $ k \in\dbc{K}$, let $\byy_k \brc{0} =\boldsymbol{0}$. For $ m \in \dbc{M}$, set 
\begin{subequations}
\begin{align}
\bww_m\brc{1} &= \argmin_{\bvv\in\setW} \left. c\brc{\bvv} \right. \\
\mR_m^{\rmw}\brc{1} &= \left[\gredSq{ } c\brc{\bww_m\brc{1}}\right]^{-1}
\end{align}
\end{subequations}
%$\bhxx_n\brc{1}$ and $\mR_n^{\rmx}\brc{1}$ for $ n\in[N]$ to their initial conditions. \vspace*{2mm}
\While\NoDo $1\leq t < T$\vspace*{1mm}
\For\NoDo $k \in\dbc{K}$ and $m \in\dbc{M}$ update
\begin{subequations}
\begin{align}
\mR_k^{\rmv} \brc{t} &= \sum_{m=1}^M \mQ_{mk} \mR_m^{\rmw} \brc{t} \mQ_{mk}^\trp \label{eq:A1} \\
%\bhzz_k \brc{t} &=\sum_{m=1}^M \mQ_{m k} \bww_m \brc{t} \label{eq:A2} \\
\bvv_k \brc{t} &= \sum_{m=1}^M \mQ_{m k} \bww_m \brc{t} -\mR_k^{\rmv} \brc{t} \byy_k \brc{t-1} \label{eq:A3} \\
\byy_k \brc{t} &=\bgg_{\out} \brc{ \bvv_k \brc{t} ,\bss_k,\mR_k^{\rmv} \brc{t} } \label{eq:A4} \\
\mR_k^{\rmy} \brc{t} &=-\gred{\bvv} \bgg_{\out} \brc{ \bvv_k \brc{t} ,\bss_k,\mR_k^{\rmv} \brc{t} }\label{eq:A5}\\
\mR_m^{\rmu} \brc{t} &= \left[ \sum_{k=1}^K \mQ_{m k}^\trp \mR_{k}^{\rmy} \brc{t} \mQ_{m k}\right]^{-1} \label{eq:A6} \\
\buu_m \brc{t} &=\bww_m \brc{t} + \mR_m^{\rmu} \brc{t}\left[ \sum_{k=1}^K \mQ_{m k}^\trp \byy_k \brc{t} \right] \label{eq:A7} \\
\bww_m \brc{t+1} &=\bgg_{\In} \brc{ \buu_m \brc{t} ,\mR_m^{\rmu} \brc{t} } \label{eq:A8} \\
\mR_n^{\rmw} \brc{t+1} &= \left[\gred{\buu} \bgg_{\In} \brc{ \buu_m \brc{t} ,\mR_m^{\rmu} \brc{t} } \right] \mR_m^{\rmu} \brc{t} \label{eq:A9}
\end{align}
\end{subequations}
\EndFor
\EndWhile \vspace*{1mm}
\Out $\bww_m \brc{T}$ for $m \in \dbc{M}$.
\end{algorithmic}
\end{algorithm}
%Following the results of \cite{bereyhi2018precoding}, it is straightforward to show that~the~precoding problem in \eqref{eq:GLSE} is iteratively solved via Algorithm~\ref{A-GAMP}. 
In this algorithm, $\bw$ is constructed after $T$ iterations. Here,
\begin{itemize}
%\item $t$ and $T$ denote the iteration counter and  the maximum number of iterations, respectively.
\item $\bww_m\brc{t}$ and $\bss_k$ are the augmented forms of the~$m$-th \ac{rf} symbol in iteration $t$, i.e. $w_m\brc{t}$, and $s_k$, respectively. %; for example $\bxx_n=[\re{x_n}, \img{x_n}]^\trp$.\vspace*{1mm}
\item $\mR^\rmw_m \brc{t}$, $\mR^\rmv_k \brc{t}$, $\mR^\rmy_k \brc{t}$ and $\mR^\rmu_m \brc{t}$ are two-dimensional real square matrices, and $\mQ_{m k}$ is defined as
\begin{align}
\mQ_{m k} \coloneqq 
\begin{bmatrix}
\Re\set{h_{mk}} &-\Im\set{h_{mk}}\\
\Im\set{h_{mk}} &\hphantom{-}\Re\set{h_{mk}}
\end{bmatrix}
\end{align}
where $h_{m k}$ is the entry $(m,k)$ of $\mH$.\vspace*{1mm}
\item $\bgg_{\out}\left(\cdot \right)$ is the output thresholder given by
\begin{align}
\bgg_{\out}\left(\bvv,\bss,\mR\right) &\coloneqq \gred{\bvv} \min_{\bzz\in\setC} \left. G_{\out} \brc{\bzz,\bvv,\bss,\mR} \right. \label{eq:g_out}
\end{align}
where $G_{\out} \brc{\cdot}$ is
\begin{align}
G_\out \brc{\bzz,\bvv,\bss,\mR} =  q \brc{\bzz - \bvv , \mR} +  \norm{\bzz-\sqrt{\rho} \left. \bss \right. }^2
\end{align}
with $q \brc{\cdot}$ being $q \brc{\bxx , \mR} =  \brc{\bxx^\trp  \mR^{-1} \bxx } / 2$.
\item $\bgg_{\In}\left(\cdot \right)$ is the input thresholding function and reads
\begin{align}
\bgg_{\In}\left( \buu,\mR\right) \coloneqq \argmin_{\bww\in\setW} G_{\In} \brc{\bww,\buu, \mR} \label{eq:g_In}
\end{align}
where $G_{\In} \brc{\cdot}$ is given by
\begin{align}
G_{\In} \brc{\bww , \buu,\mR} =  q \brc{\buu - \bww , \mR} + c\brc{\bww}. \label{eq:E_in}
\end{align}
\end{itemize}

\subsection{Projection on the Precoding Support}
\label{sec:Pro_back}
To project the precoded \ac{rf} signal $\bw$ back on the precoding support, we follow the approach proposed in Section~\ref{sec:projection}.~To this end, transmit entry $x_m$ for $m\in\dbc{M}$ is calculated via~\eqref{eq:RevRF3} for some $\theta$ and regularization term $r\brc{\cdot}$. To tune $\theta$, we note %that
\begin{enumerate}
\item As $\theta \downarrow 0$, \eqref{eq:RevRF3} determines the minimizer of $r\brc\cdot$ over~the set of points $u$ at which $w_m = f_\rf\brc{u}$. This tuning~is~sui- table for \ac{rf} conversion functions which accurately model the input-output characteristic of the \ac{rf} chains,~i.e. when $\epsilon \downarrow 0$ in Remark \ref{remark:1}.
\item For $\theta \uparrow \infty$, \eqref{eq:RevRF3} finds the minimizer of $r\brc\cdot$ over $\setC$. Such a setting corresponds to \ac{rf} conversion models with high error,~i.e. when $\epsilon$ is significantly large.
\end{enumerate}
As a result, $\theta$ is tuned such that it monotonically increases with $\epsilon$, where $\epsilon$ is error of the analytic model given by $f_\rf \brc\cdot$.

\section{Numerical Investigations}
In this section, we investigate the performance~of~the~proposed algorithm through numerical simulations. For this aim, we first specify the configuration which is being simulated.%, and define metrics which quantify the performance of the system.

\subsection{System Configuration}
We consider the case in which \ac{iid} zero-mean and unit-variance Gaussian data symbols $s_1,\ldots,s_K$ are to be transmitted over the downlink channel. It is assumed that the channel experiences \ac{iid} Rayleigh fading. This means that the entries of $\mH$ are \ac{iid} Gaussian with zero-mean and variance $1/M$. The characteristics of each system component are illustrated in the sequel.

\subsubsection{RF chains}
The conversion function of an \ac{rf} chain~is~assumed to be fully described via its \ac{pa}, and the input-output characteristics of the \ac{pa} is represented by the amplitude-to-amplitude and amplitude-to-phase model which reads\vspace{-1mm}
\begin{align}
f_\rf\brc{x} = f_{\rm A} \brc{\abs{x}} \exp\set{ \rmj f_{\Phi} \brc{\abs{x}} } \left. \frac{x}{\abs{x}} \right. .
\end{align}
In this model, $f_{\rm A} \brc{\cdot}$ is the amplitude-to-amplitude conversion function which specifies the amplitude of the \ac{rf} signal at the output of the \ac{pa}. $f_{\Phi} \brc{\cdot}$ further determines the nonlinear phase shift at the output which is a function of the input amplitude.

Well-known analytic formulations for $f_{\rm A} \brc{\cdot}$ and $f_{\Phi} \brc{\cdot}$ are given by Saleh's model \cite{saleh1981frequency,o2009new} in which\vspace{-1mm}
%\begin{subequations}
\begin{align}
f_{\rm A} \brc{ \omega } = \frac{\alpha_{\rm A} \omega }{ 1+ \beta_{\rm A} \omega^2},\qquad
f_{\Phi} \brc{ \omega } = \frac{\alpha_{\Phi} \omega^2 }{ 1+ \beta_{\Phi} \omega^2}.
\end{align}
%\end{subequations}
Here, $\brc{\alpha_{\rm A}, \beta_{\rm A}}$ and $\brc{\alpha_{\Phi} , \beta_{\Phi}}$ are non-negative scalars which are determined for a specific \ac{pa} numerically via the method~of least-squares. The model is assumed to have average error~$\epsilon$. This means that for the true output symbol $w$, we have\vspace{-1mm}
\begin{align}
\Ex{\abs{ w-f_\rf\brc{x} }^2}{x} \leq \epsilon
\end{align}
where $x$ is the true input symbol, and $\Ex{\cdot}{x}$ averages over~all possible realizations of $x$.

\subsubsection{Precoder}
From the example in Fig.~\ref{fig:1}, we know that~the \ac{pa}'s output is saturated at some level. This means that the \ac{rf} transmit entries always satisfy $\abs{w_m}^2 \leq P_\out$ for some power $P_\out$ which is specified for each \ac{pa}. As a result, we consider the \ac{rf} constellation set as %$\setW$ to be
%\begin{align*}
$\setW = \set{ w\in \setC : \abs{w} \leq \sqrt{P_\out} }$. %
%\end{align*}

The precoding support $\setX$ is further set to $\setX=\setC$. Following the discussions in Section~\ref{sec:Pro_back}, we need to set $\theta$~in~\eqref{eq:RevRF3}~monotonically increasing in $\epsilon$ for backward projection of the \ac{rf} transmit signal on $\setX$. We consider the simple choice of $\theta = \epsilon$ and set $r\brc{u} = \abs{u}^2$. 

For the sake of simplicity, we assume full transmit complexity, i.e. $L=M$, with limited average transmit power; hence, $c\brc{\bw} = \lambda \norm{\bw}^2$ for some $\lambda$. Nevertheless, scenarios~with~partially active arrays, i.e. $L\leq M$, are straightforwardly~addressed by modifying $c\brc\cdot$; see discussions in \cite{bereyhi2018glse}.

\subsection{Performance Metrics}
Following the discussions in Section~\ref{sec:RLS}, we know that the \ac{rss} defined in Definition~\ref{def:RSS} quantifies the performance of the precoder, effectively. We hence define the performance metric with respect to the \ac{rss}. To this end, let $\bw \in \setW^{M}$~be~the~signal precoded directly at the \ac{rf} stage via \eqref{eq:GLSE}. The transmit signal $\bx\in\setC^M$ is calculated entry-wise from $\bw$ using \eqref{eq:RevRF3}. The true \ac{rf} signal is then given by $\tilde{\bw} = f_\rf\brc{\bx}$ which is in general different\footnote{Note that \eqref{eq:RevRF3} solves exactly $\bw = f_\rf\brc{\bx}$ only when $\theta = 0$.} from $\bw$. In this case, the average \ac{rss} predicted by the \ac{rf}-stage \ac{glse} precoder is
\begin{align*}
D\brc{\rho} = \frac{1}{K}\norm{ \mH^\trp \bw  - \left. \sqrt{\rho} \right. \bs  }^2
\end{align*}
for the given scaling factor $\rho$. However, the average \ac{rss} which is achieved in practice is
\begin{align*}
\tilde{D} \brc{\rho} = \frac{1}{K}\norm{ \mH^\trp \tilde\bw  - \left. \sqrt{\rho} \right. \bs  }^2.
\end{align*}
$D \brc{\rho}$ and $\tilde{D} \brc{\rho}$ in general address the average distortion~imposed by the multiuser interference at each UT. For effective design of the precoder and small $\epsilon$, we have $\tilde{D} \brc{\rho}\approx D\brc{\rho}$.

\subsection{Numerical Results}
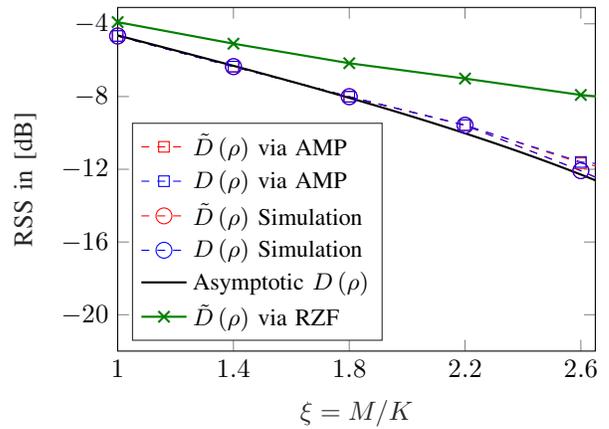
\begin{figure}[t]
\centering
\definecolor{mycolor1}{rgb}{0.00000,0.49804,0.00000}%
\begin{tikzpicture}

\begin{axis}[%
width=2.5in,
height=1.8in,
at={(1.976in,1.234in)},
scale only axis,
xmin=1,
xmax=2.65,
xtick={1,1.4,1.8,2.2,2.6},
xticklabels={{$1$},{$1.4$},{$1.8$},{$2.2$},{$2.6$}},
xlabel style={font=\color{white!15!black}},
xlabel={$\xi= M/K$},
ymin=-22,
ymax=-3.1,
ytick={-20,-16,-12,-8,-4},
yticklabels={{$-20$},{$-16$},{$-12$},{$-8$},{$-4$}},
ylabel style={font=\color{white!15!black}},
ylabel={RSS in [dB]},
axis background/.style={fill=white},
legend style={at={(0.03,0.03)}, anchor=south west, legend cell align=left, align=left, draw=white!15!black}
]
\addplot [color=red, dashed, mark size=2pt, mark=square, mark options={solid, red}]
  table[row sep=crcr]{%
1	-4.66847815138557\\
1.4	-6.35292348286973\\
1.8	-8.0065518936445\\
2.2	-9.57106627696627\\
2.6	-11.6632981916836\\
3	-12.8332953029041\\
};
\addlegendentry{\small{$\tilde{D} \brc{\rho}$ via AMP}}

\addplot [color=blue, dashed, mark size=2pt, mark=square, mark options={solid, blue}]
  table[row sep=crcr]{%
1	-4.66976351169748\\
1.4	-6.35512923720011\\
1.8	-8.00979947597604\\
2.2	-9.5763766620436\\
2.6	-11.596121655596\\
3	-12.2668450200501\\
};
\addlegendentry{\small{${D} \brc{\rho}$ via AMP}}

\addplot [color=red, dashed, mark size=3.0pt, mark=o, mark options={solid, red}]
  table[row sep=crcr]{%
1	-4.6684630104673\\
1.4	-6.35289531194929\\
1.8	-8.00665945646893\\
2.2	-9.57676614436639\\
2.6	-12.0724190628725\\
3	-14.9208761472586\\
};
\addlegendentry{\small{$\tilde{D} \brc{\rho}$ Simulation}}

\addplot [color=blue, dashed, mark size=3.0pt, mark=o, mark options={solid, blue}]
  table[row sep=crcr]{%
1	-4.66974370373582\\
1.4	-6.35507481174546\\
1.8	-8.0098233580375\\
2.2	-9.58190534056112\\
2.6	-12.0801106192424\\
3	-14.9354338941549\\
};
\addlegendentry{\small{$D \brc{\rho}$ Simulation}}

\addplot [color=black, line width=.8pt]
  table[row sep=crcr]{%
1	-4.63571323918141\\
1.05	-4.84331491693553\\
1.1	-5.05093345984616\\
1.15	-5.25879713365047\\
1.2	-5.46712066981614\\
1.25	-5.67610801041749\\
1.3	-5.88595462550153\\
1.35	-6.09684949866049\\
1.4	-6.30897685475866\\
1.45	-6.52251769044443\\
1.5	-6.73765115006052\\
1.55	-6.95455578709702\\
1.6	-7.17341074095959\\
1.65	-7.39439685344525\\
1.7	-7.61769775076768\\
1.75	-7.84350090601883\\
1.8	-8.0719987049436\\
1.85	-8.30338952956788\\
1.9	-8.53787887617581\\
1.95	-8.7756805251358\\
2	-9.01701777986792\\
2.05	-9.26211936043885\\
2.1	-9.51124185863334\\
2.15	-9.76464073577722\\
2.2	-10.0225918285695\\
2.25	-10.2853884416479\\
2.3	-10.553343467506\\
2.35	-10.826791762534\\
2.4	-11.1060928251317\\
2.45	-11.3916338305017\\
2.5	-11.6838330875025\\
2.55	-11.983143996362\\
2.6	-12.2900596028882\\
2.65	-12.6051178659854\\
2.7	-12.928907782137\\
2.75	-13.2620765447783\\
2.8	-13.6053379604917\\
2.85	-13.9594824008829\\
2.9	-14.3253886436308\\
2.95	-14.7040380539733\\
3	-15.0965316897305\\
};
\addlegendentry{\small{Asymptotic $D \brc{\rho}$}}

\addplot [color=mycolor1,line width=.8pt, mark size=3.0pt, mark=x, mark options={solid, mycolor1}]
  table[row sep=crcr]{%
1	-3.90549163015619\\
1.4	-5.09190416830498\\
1.8	-6.16684457185917\\
2.2	-7.01083714157389\\
2.6	-7.91187077041099\\
3	-8.51422142104351\\
};
\addlegendentry{\small{$\tilde{D} \brc{\rho}$ via RZF}}

\end{axis}
\end{tikzpicture}%
\caption{Average\ac{rss} vs. per-user number of antennas for $\rho =1$. The \ac{papr} of the \ac{rf} signal is set to $\log \mathrm{PAPR} = 5$ dB.\vspace*{-4mm}}
\label{fig:2}
\end{figure}

Fig.~\ref{fig:2} plots $D \brc{\rho}$ and $\tilde{D} \brc{\rho}$ against the number of transmit antennas per user, i.e. $\xi = M/K$, for $\rho = 1$. The transmit array size is set to $M=64$, and the parameters of the \ac{pa} read
\begin{align*}
\brc{\alpha_{\rm A} , \beta_{\rm A} }= \brc{2.159 , 1.152}, \qquad \brc{\alpha_{\Phi}, \beta_{\Phi}}= \brc{4.003,9.104}
\end{align*}
with $\epsilon = 0.05$. Considering the dynamic range of the \ac{pa}, the peak output power on the \ac{rf} stage is set to $P_{\rm out}=1$. 

To compare the results with the benchmark, $\lambda$ is tuned, such that the \ac{papr} of the \ac{rf} transmit signal is $\log \mathrm{PAPR} = 5$ dB. The results are given for~Algorithm~\ref{A-GAMP}, as well as directly solving \eqref{eq:GLSE} via CVX \cite{cvx,gb08}.

It is observed that $D \brc{\rho}$ closely matches $\tilde{D} \brc{\rho}$~which~indicates the efficiency of the backward projection. The figure further depicts the accuracy of Algorithm~\ref{A-GAMP}, as its results~are tightly consistent with the direct simulations. For~sake~of~comparison, we sketch two other plots: The first plot shows the asymptotic value of $D \brc{\rho}$ derived in \cite{bereyhi2018glse}. This plot is closely tracked by the finite-dimension simulations. The second plot shows $\tilde{D} \brc{\rho}$ achieved via \ac{rzf} precoding \cite{peel2005vector}. The \ac{rzf} precoder in this case is tuned,~such~that the output \ac{papr} remains $\log \mathrm{PAPR} = 5$ dB. As the plot demonstrates, \ac{rzf} precoding exhibits degraded performance. This is a result of unwanted distortion imposed by nonlinear characteristics of the \ac{rf} chains. Due to the page limit, further numerical investigations are skipped and will be given in the extended version of the manuscript.

\section{Conclusions}
The proposed precoding scheme for massive \ac{mimo} settings with nonlinear front-ends utilizes the \ac{rls} method to jointly invert the channel and compensate distortions caused by nonlinear \ac{rf} chains. An \ac{amp}-based algorithm implements the proposed scheme with low complexity. Numerical investigations show performance enhancement compared to the classic precoding techniques. Although the main focus of this paper was on the \ac{pa}, the results are straightforwardly extended to other non-ideal transceiver components. %Further investigations~will~be~given in the extended manuscript.

\bibliography{ref}
\bibliographystyle{IEEEtran}
\end{document}